\documentclass[11pt]{article}

\usepackage{amssymb,amsmath}
\usepackage[dvips]{graphicx}
\usepackage{epsfig}
\usepackage{cite}

\def\laq{~\raise 0.4ex\hbox{$<$}\kern -0.8em\lower 0.62
ex\hbox{$\sim$}~}
\def\gaq{~\raise 0.4ex\hbox{$>$}\kern -0.7em\lower 0.62
ex\hbox{$\sim$}~}
\numberwithin{equation}{section}

\def\be{\begin{equation}}
\def\ee{\end{equation}}
\def\bea{\begin{eqnarray}}
\def\eea{\end{eqnarray}}
\newcommand{\nn}{\nonumber}
\newcommand{\de}{\partial}

\def \l {\lambda}
\def \L {\Lambda}
\def \d {\delta}

\def \b {\beta}

\def \g {\gamma}
\def \s {\sigma}

\def \6 {{}^{(6)}}
\def \5 {{}^{(5)}}
\def \4 {{}^{(4)}}

\def \rarr {\rightarrow}

\begin{document}

\begin{titlepage}

\begin{flushright}
BA-TH 620-09
\end{flushright}

\vspace*{1.5 cm}

\begin{center}

\Huge
{Induced cosmology on a codimension-2 brane in a conical bulk}

\vspace{1cm}

\large{G. De Risi}

\normalsize
\vspace{.2in}
{\sl Dipartimento di Fisica, Universit\`a degli Studi di Bari, \\
and} \\
{\sl Istituto Nazionale di Fisica Nucleare, Sezione di Bari\\
Via G. Amendola 173, 70126 Bari, Italy}

\vspace*{1.5cm}

\begin{abstract}
We study the cosmology of a 5-dimensional brane, which represents a regularization of
a 4-dimensional codimension-2 brane, embedded in a conical bulk. The brane is assumed
to be tensional, and to contain a curvature term. Cosmology is obtained by letting the brane
move trough the bulk, and implementing dynamical junction conditions. Our results shows
that, with suitable choices of the parameters, the resulting cosmological dynamics mimics
fairly well standard 4-dimensional cosmology.

\end{abstract}
\end{center}

\end{titlepage}

\newpage

\section{Introduction}
\label{intro}

For more than ten years now, the appealing idea that our universe is a ``brane'' embedded in
a higher dimensional bulk has been extensively explored, and widely accepted as a serious
alternative to standard 4-dimensional cosmological models to solve the longstanding problems
that prevents the complete understanding of our universe.

Codimension-1 branes, since the
seminal paper by Randall and Sundrum \cite{Randall:1999vf}, have been proved to be
viable in allowing both a low-energy limit that mimics Newtonian gravity, and a compelling
cosmological dynamics that can accommodate for inflation,
match cosmological observations with respect to scalar and tensor perturbations,
as well as opening novel possibilities for addressing both the initial singularity and
the late time acceleration problem (for a review on these ad other aspects of Randall-Sundrum
cosmology, see, for example, \cite{Langlois:2002bb,Maartens:2003tw}).

Codimension-2 branes are, in some sense, even more interesting, but, unfortunately, also much less
``feasible''. In fact, the celebrated ADD mechanism \cite{ArkaniHamed:1998rs} to address the
hierarchy problem is viable only with two (or more) extra dimension. In addition,
a proposal has been put forward \cite{Carroll:2003db} that the vacuum energy of a codimension-2
brane can be ``off-loaded'' in the bulk. Brane geometry would stay flat, and the energy will
generate a deficit angle in the codimension-2 bulk (thus generating a conical singularity at the
origin). This idea could explain the absence of a cosmological constant originated
from vacuum fluctuations in quantum field theory. In this approach, the current small
value observed for the cosmological constant should be generated by some different mechanism,
which can be provided by a generalization of this proposal, the so-called Supersymmetric Large
Extra Dimension model (SLED) \cite{Aghababaie:2003wz}.
The idea is that supersymmetry-breaking on the brane at high energy, which do not generate a vacuum
energy because of the self-tuning property, induces a supersymmetry-breaking
scale in the bulk at a much lower energy which depends on the size of the extra
dimensions. The desired size to solve the hierarchy problem within the ADD framework
provides an order of magnitude for the bulk SUSY breaking scale that generates, back
on the brane, a cosmological constant with the correct order of magnitude.
However, this proposal, even in the supersymmetric extension, has met severe criticism
\cite{Garriga:2004tq}, because the self-tuning property rely on a tuning of
the magnetic flux that stabilize the extra dimensions, which can
not be kept stable under a phase transition on the brane.

Even worse, it is not possible to accommodate on a codimension-2 brane any kind
of energy-momentum tensor different from the pure tension \cite{Vinet:2004bk}. So,
to study low energy limit, and possibly cosmology, one has to implement some kind
of regularization of the 4-dimensional brane. Several regularization has been proposed
\cite{Peloso:2006cq,Kaloper:2007ap,Burgess:2006ds,Burgess:2007vi,Kobayashi:2007qe,Burgess:2008yx},
in which linear analysis show that weak gravity has the tensor structure of  general relativity,
but with the presence of some long-range modulus which should disappear from the spectrum
at the nonlinear level. The key point then, is how to describe a non-trivial dynamics
of the regularized brane, and eventually how to derive from this description a viable
4-dimensional cosmology. This is, as one can imagine, a formidable task, because it would
require to handle with the complete 6-dimensional dynamics. Some attempts have been made
\cite{Papantonopoulos:2007fk,Minamitsuji:2007fx,Copeland:2007ur,Kobayashi:2007hf,Kobayashi:2008bm}
using approximation to tackle the problem.

In this paper we study cosmology of a regularized brane in a 6-dimensional conical bulk.
The dynamical equations are obtained by letting the regularized brane move trough
the bulk and implementing dynamical junction conditions \cite{Kraus:1999it,Kehagias:1999vr}.
The model we present can be seen as a warped generalization of the conical codimension-2 brane of
\cite{Kaloper:2007ap}, but, since we are interested in cosmology on the brane, we are {\it not}
in need of explicitly adding an axion field to recover a 4-dimensional tensional brane
(a similar contribution can be obtained by a particular choice of the parameters
of the energy-momentum tensor of the brane, see section \ref{junction_sec}).
Furthermore, the present model differs from the standard ``rugby-ball'' regularization
because the radial extra-dimension is non-compact, and the space-time is ``capped''
only in the inner side of the bulk. This could be a great advantage, because
the dynamical behavior of the model should have some peculiar features
that cannot be obtained by a low energy KK reduction of a 6-dimensional theory, but could as well
leads to completely unacceptable behavior in the regime in which the wrapping of the
brane becomes too large. We simply avoid this problem assuming that the brane movement
is always close enough to the inner cap (so that even the ``late intrinsic time'' regime
must be assumed to respect this limit); this condition can be achieved by tuning conveniently the
deficit angle to a near-critical value (which is the same request to obtain a correct tensorial structure in the
linear approximation \cite{Kaloper:2007ap}). With this precaution, and under some assumptions
described below, we find that unlike other models studied with similar techniques,
4-dimensional induced cosmology can mimics fairly well the standard cosmological model,
with an initial singularity and accelerated expansion at late time. The latter is driven by
an effective cosmological constant given by a particular combination of the bulk cosmological constant and
the brane tension.

The paper is organized as follows: In section \ref{static} we present the
static solution and the set up of the 5-dimensional brane. Then, in section \ref{junction_sec}
we derive the cosmological equation on the brane with the junction conditions.
These equations are studied in section \ref{cosmol_sect}, with some
assumptions that allows us to solve equations analytically and, where not possible, numerically.
Finally, in section \ref{comm_concl} we comments on the results obtained in the previous section
and draw our conclusions.

\section{The static solution}
\label{static}

We consider a 6-dimensional space with cosmological constant, in which a 5-dimensional brane
is embedded. We assume the brane to be tensional, and to have a curvature term,
as well as matter, on it. The action that describes the model is
\be
S = \int d^6 x \sqrt{-g} \left( \frac{M^4}{2}R - \L_6 \right) - \int d^5 \xi \sqrt{\g}
\left( \frac{M_5^3}{2} ~ \5 R + \mathcal{L}_{brane} \right).
\label{action}
\ee
Capital latin indices ($A$, $B$ \ldots) run from $0$ to $5$ (so refer to bulk objects), while greek
indices ($\mu$, $\nu$, \ldots) run from $0$ to $4$ (so refer to brane objects). The brane
intrinsic metric is $\g_{\mu \nu}$. The equations of motion following from
this action, far from the brane, are:
\be
G_{AB} = -\L g_{AB},
\label{vac_eq}
\ee
with $\L = \L_6/M^4$. We do not consider the singular contribution coming from the brane
in (\ref{vac_eq}), because we will take it into account later via the junction conditions.
We seek for a metric which has a flat 4-dimensional submanifold, to be identified with our universe,
and the two remaining dimensions having the geometry of a cone \cite{Kaloper:2007ap}.
The metric is:
\be
ds_6^2 = R^2 ds_4^2 + \left[ \frac{\L}{10}\left( R^2 - \frac{\mu^5}{R^3}\right) \right]^{-1} dR^2 +
\beta^2 \ell^2 \left( R^2 - \frac{\mu^5}{R^3}\right) d\chi^2.
\label{vac_metric}
\ee
with $ds_4^2 = \eta_{\mu \nu} dx^{\mu} dx^{\nu}$. This metric can have a conical singularity
in $R = \mu = R_h$, in addition to the ``true'' singularity in $R = 0$.
Upon defining the new coordinate $\rho$ by
\be
d \rho^2 = \left[ \frac{\L \mu}{2} \left( R - \mu\right)\right]^{-1} dR^2
\label{rho}
\ee
the extradimensional part of the metric, close to the horizon, can be approximated by:
\be
ds_2^2 \simeq d \rho^2 + \frac{5}{8}\b^2 \ell^2 \mu^2 \L \rho^2 d\chi^2
\label{cone_metric}
\ee
thus we see that the metric is regular close to the horizon if
\be
\b^2 \ell^2 = \left( \frac{5}{8}\mu^2 \L \right)^{-1}
\label{flat_par}
\ee

The space-time of our model consists of two manifolds described by the metric (\ref{vac_metric}),
joined together at the radial position of the brane $R = R_b$:
\bea
&& ds_{6,in}^2 = z_i^2 ds_4^2 + \left[ \frac{\L_i}{10}\left( z_i^2 -
\frac{\mu_i^5}{z_i^3}\right) \right]^{-1} dR^2 + \beta_i^2 \ell_i^2 \left( z_i^2 -
\frac{\mu_i^5}{z_i^3}\right) d\chi^2, \nn \\
&& ds_{6,out}^2 = z_o^2 ds_4^2 + \left[ \frac{\L_o}{10}\left( z_o^2 -
\frac{\mu_o^5}{z_o^3} \right) \right]^{-1} dR^2 + \beta_o^2 \ell_o^2
\left( z_o^2 - \frac{\mu_o^5}{z_o^3} \right) d\chi^2.
\label{in-out_metric}
\eea
where $z = R/R_0 + C$. Continuity along the brane directions requires $z_i = z_o$, while
continuity along the compact extradimensional direction $\chi$ gives
\be
\mu_i = \mu_o,~~~~~ \b_i^2 \ell_i^2 = \b_o^2 \ell_o^2 = \b^2 \ell^2.
\label{cont_par}
\ee
Then we can, without any loss
of generality, set the integration constants $C = \mu$, so that the space-time ``begins''
at $R = 0$, and $R_0 = 1$ (a different choice would just end up in a rescaling of the
cosmological constant). So the two parts of the space-time differ only because of
the different values of the cosmological constants.
Having in mind to study mirage cosmology on the brane by allowing it to move trough the
radial direction, we can impose the relation (\ref{flat_par}) only for the $in$ part of the
space-time, so that it ends smoothly at the position of the horizon. On the other side,
the $out$ part of the space-time is allowed to have a deficit angle $1-b$, so that the
codimension metric written using the variable $\rho$ of (\ref{rho}) reads:
\be
ds^2_{2,out} = d\rho^2 + (1-b)^2 \rho^2 d\chi^2.
\label{cone_metric2}
\ee
This fixes the relation between the cosmological constants of the $in$ and $out$ parts of
the space-time to be:
\be
\L_i = \frac{\L_o}{(1-b)^2}
\label{cc_rel}
\ee

\section{Cosmological equations on the brane}
\label{junction_sec}

The presence of matter on the brane, and the movement of the brane itself across the
extra-dimension would in principle modify the bulk geometry. Solving exactly this problem
would be extremely complicated, so we assume that cosmology is induced on the brane
by implementing time-dependent Israel junction condition, while the bulk
is not modified by the brane movement \cite{Papantonopoulos:2007fk,Minamitsuji:2007fx}.
Let us assume then that the brane position is $R_b \equiv a(\tau)$. The brane embedding is thus
described by the relation between the bulk and the brane coordinates
$\xi^a = (\tau,{\bf x},\chi)$:
\be
t = t(\tau),~~~~~~ R = a(\tau)
\label{embedding}
\ee
the other relations being just identities. So the tangent vectors are trivial, except the
timelike one $u_t^A$, which reads, in the coordinate system we are using:
\be
u_t^A = \left(\dot{t},{\bf 0},\dot{a},0 \right)
\label{t_vect}
\ee
where dot indicates derivative with respect to $\tau$. The normal vector $n^A$ can be expressed as
\be
n^A = \left(n^t,{\bf 0},n^R,0 \right).
\label{n_vect}
\ee
By using the orthogonality conditions $g_{AB}u^A u^B = 1$, $g_{AB}n^A n^B = -1$, $g_{AB}u^A n^B = 0$
we can express all the unknown functions in terms of the scale factor $a(\tau)$. Of course, the
orthogonality conditions are different on the two sides of the brane, because of the difference
between the $in$ and $out$ metric. We have
\bea
\dot{t} = \frac{\sqrt{\dot{a}^2 + z^2 f_i(z)}}{z\sqrt{f_i(z)}}, &~~~~~
n^R = -\sqrt{\dot{a}^2 + z^2 f_i(z)}, &~~~~~ n^t = -\frac{\dot{a}}{z^2\sqrt{f_i(z)}} \nn \\
\dot{t} = \frac{\sqrt{\dot{a}^2 + z^2 f_o(z)}}{z\sqrt{f_o(z)}}, &~~~~~
n^R = \sqrt{\dot{a}^2 + z^2 f_o(z)}, &~~~~~ n^t = \frac{\dot{a}}{z^2\sqrt{f_o(z)}}
\label{vec_values}
\eea
with
\be
f_{i/o}(z) = \frac{\L_{i/o}}{10}\left(1 - \frac{\mu^5}{z^5} \right).
\label{f}
\ee
the difference is due to the different value of the cosmological constants and to the different
orientation of the normal unit vector in the two branches. With the normal unit vector
in hands, we can calculate the induced metric $h_{AB} = g_{AB} - n_A n_B$, and
consequently the intrinsic line element on the brane:
\be
ds^2_5 = -d\tau^2 + z^2(\tau) {\bf dx}^2 + \b^2 \ell^2 \left( z^2(\tau) -
\frac{\mu^5}{z^3(\tau)} \right) d\chi^2
\label{intr_metric}
\ee
which is the same on both side of the by means of (\ref{cont_par}), as expected.
Then we can evaluate the extrinsic curvature $K_{AB} = h_A^{~~C} \nabla_C n_B$,
noting that derivatives w.r.t. the bulk variables are expressed in terms
of derivatives w.r.t. brane variables via the chain relation \cite{Papantonopoulos:2007fk}
\be
\de_A = g_{AB} \de_\mu x^B \g^{\mu \nu} \de_\nu.
\label{chain_formula}
\ee
We find
\bea
K_{tt} &=& \pm z \sqrt{\dot{a}^2 + z^2 f_{i/o}} \left[ \frac{\ddot{a}}{z f_{i/o}} -
\frac{\dot{a}f'_{i/o}}{2z^2 f^2_{i/o}} + 1 \right] \nn \\
K_{ij} &=& \mp z \sqrt{\dot{a}^2 + z^2 f_{i/o}} \d_{ij} \nn \\
K_{RR} &=& \frac{\dot{a}^2}{z^4 f_{i/o}\left( \dot{a}^2 + f_{i/o} \right)} K_{tt} \nn \\
K_{Rt} &=& - \frac{\dot{a}}{z^2 \sqrt{f_{i/o}\left( \dot{a}^2 + f_{i/o}  \right)}} K_{tt} \nn \\
K_{\chi \chi} &=& \mp \frac{8\left( 2zf_{i/o} + z^2 f'_{i/o} \right)}{\mu^2 \L_{i/o}^2}
\sqrt{\dot{a}^2 + z^2 f_{i/o}}
\label{K}
\eea
where the upper sign refers to the $in$ side of the brane, and the prime indicates derivative
w.r.t. $z$. Equations of motion on the brane are obtained by equating the discontinuity
of the projected extrinsic curvature across the brane with the energy-momentum contribution
on the brane (which, in our case, includes the curvature term):
\be
\left[ K_{\mu \nu} \right] - \left[ K \right] \gamma_{\mu \nu} = \frac{1}{M^4}
\left( T_{\mu \nu} - M_5^3  G_{\mu \nu}\right)
\label{brane_eom}
\ee
where $[x]$ stands for $x_o - x_i$, $K_{\mu \nu} = u^A_{~~\mu} u^B_{~~\nu} K_{AB}$ and
$K$ is its trace, $G_{\mu \nu}$ is the intrinsic Einstein tensor as calculated
from the intrinsic metric $\g_{\mu \nu}$. The brane energy-momentum tensor needs to be
specified. We assume, in addition to the tension contribution, a
``perfect fluid-like'' form, which is compatible with the symmetry of the space-time:
\bea
T_\mu^{~~\nu} &=& -\l \eta_\mu ^{~~\nu} + {}^{(p.f.)}T_\mu^{~~\nu} \nn \\
{}^{(p.f.)}T_\mu^{~~\nu} &=& {\rm diag}\left( -\rho,p,p,p,P\right).
\label{em_tensor}
\eea

Let us stress that with the particular choice of the form of the e.m.
tensor (\ref{em_tensor}), the cosmological constant on the brane is taken into account separately,
so that we can impose $w > -1$. Of course this is restrictive, since the symmetry of the
space-time allows an energy momentum tensor with $\rho = -p = \lambda$ and $P=-\lambda'$,
thus having a sort of ``ring'' cosmological constant that can differs from the 4-dimensional
one. We will comment more on this in the next section.

Finally, substituting (\ref{brane_eom}) in  (\ref{em_tensor}), after some algebraic manipulation,
the equations of motion read (from now on, we drop the subscript $i/o$, assuming that all quantities
are intended to be in the $out$ side, and use (\ref{cc_rel}) to express appropriately the
correspondent $in$ objects):
\bea
&& \sqrt{H^2 + f} + \sqrt{H^2 + \s^2 f} = \frac{2}{M^4}\frac{1 -
\left( \frac{\mu}{z} \right)^5}{8-3\left( \frac{\mu}{z} \right)^5}
\left( \rho + \l \right) - 3 r_c \frac{4 + \left( \frac{\mu}{z}
\right)^5}{8-3\left( \frac{\mu}{z} \right)^5}H^2  \label{compeq_1} \\
&& \dot{\rho} + H\left( \frac{(4 + 3w) - 3(\frac{1}{2} + w)\left( \frac{\mu}{z} \right)^5}
{1-\left( \frac{\mu}{z} \right)^5}\rho - \frac{1 + \frac{3}{2} \left( \frac{\mu}{z} \right)^5}
{1-\left( \frac{\mu}{z} \right)^5} P \right) = 0  \label{compeq_2} \\
&& -\dot{H}\left[ \left(H^2 + f \right)^{-\frac{1}{2}} + \left(H^2 + \s^2f \right)^{-\frac{1}{2}}
\right] - \nn \\
&& -H^2 \left[ \left( 5 - \frac{1 + \frac{3}{2} \left( \frac{\mu}{z}
\right)^5}{1-\left( \frac{\mu}{z} \right)^5} \right) \left( \left(H^2 + f \right)^{-\frac{1}{2}} +
\left(H^2 + \s^2f \right)^{-\frac{1}{2}} \right) + 3 \frac{\sqrt{H^2 + \s^2 f}}
{H^2 + f} \right] - \nn \\
&& -f \left[ 4\left(H^2 + f \right)^{-\frac{1}{2}} + \s^2 \left(H^2 + \s^2f \right)^{-\frac{1}{2}}
+ 3 \frac{\sqrt{H^2 + \s^2 f}}{H^2 + f} \right] = \nn \\
&& = \frac{1}{M^4} \left( P - \l \right) + 3 r_c \left( \dot{H} + 4H^2 \right) \label{compeq_3}
\eea
with $H = \dot{z}/z$, $\s = (1-b)^{-1}$, $r_c = M_5^3/M^4$.

Eq. (\ref{compeq_1}) represents the modified Friedman equation that controls the cosmological evolution of the
5-dimensional brane. Notice, however, that $H$ is the actual 4-dimensional Hubble parameter, since it is obtained
from the scale factor that controls the dynamics of the 4-dimensional slice of the brane. Since the brane
is wrapped around the azimuthal direction, in order to obtain ``sensible'' 4-dimensional sources, we must integrate
the energy density $\rho$ over the fifth dimension \cite{Papantonopoulos:2007fk,Minamitsuji:2007fx}:
\be
\4 \rho = \int d\chi \sqrt{\g_{\chi \chi}} \rho =  2 \pi \sqrt{\g_{\chi \chi}} \rho
\label{4-D_rho}
\ee
since we assume that neither the metric nor the energy density depend on the azimuthal coordinate.
The modified Friedman eq. (\ref{compeq_1}) could be cast in a more ``conventional'' form $H^2 = f(\4 \rho)$ by
solving it for $H^2$ and inserting $\4 \rho$, but its form would be overcomplicated,
and we will show that, if one seeks for solutions only in particular regimes,
``effective'' Friedman equations will regain their simplicity.

Eq. (\ref{compeq_2}) is the conservation equation for the energy-momentum tensor (\ref{em_tensor}), in which we
have imposed an equation of state that relates only energy density and pressure, $p = w \rho$.
The symmetry of the 5-dimensional space-time leads us to include an extra-dimensional component which is undetectable
(unless one includes gauge coupling between ordinary matter and extradimensional one), but which modifies
the dynamic of the 4-dimensional universe. The presence of such a ``dark'' term is quite common in braneworld
models \cite{Binetruy:1999ut,Binetruy:1999hy} and in our model plays the crucial role of slowing down
the expansion of the energy density (in some particular regimes) thus compensating the leakage due to higher
dimensionality.
Eq. (\ref{compeq_3}) is the fifth component of the 5-dimensional junction conditions, and (by means of
the Bianchi identities) it is related to the fifth component of the conservation equation of the energy-momentum
tensor. We assume that $P$ satisfies this equation,which is then a constraint equation for the
extradimensional pressure. It is possible to imagine a more complicated scenario in which
the extradimensional pressure is not assumed to satisfy a constraint equation, but (maybe more physically)
to be related to the energy density by a general equation of state. in this case the system would look like
being not compatible, since there would be more equations than degrees of freedom. Actually, a more general
extradimensional equation of state would induce a time-dependent tilt in the azimuthal direction, which results
in a deformation of the ring shape of the brane. Studying a system like this would be very complicated, and will
probably not give an acceptable 4D cosmology because it would be very hard to get a FRW-like 4-dimensional
slicing of the 5-brane. This is the reason we assume the brane stays rigid during its movement trough the cone.

The system of eqs. (\ref{compeq_1}-\ref{compeq_3}) looks very complicated to handle. Nevertheless,
it is possible, as already anticipated, to make some assumptions that allows us to simplify it, so to
get an analytic solutions. This will be the aim of the next section.

\section{Cosmology on the brane}
\label{cosmol_sect}

It is more convenient to track back the cosmological evolution of the brane from late to early
times. Let's then first consider what happens at late times. We can guess (assuming that
the universe is expanding) that $a(\tau) \gg \mu $ so that $f$ will become just proportional
to the cosmological constant. In addition, we can assume that the energy density is negligible
with respect to the cosmological constant itself\footnote{Of course, current observations
suggest that we are actually living in a very special time in the evolution of our universe
in which matter energy density and cosmological constant are of the same order of magnitude.
We will not pursue any suggestion about the resolution of this ``coincidence'' problem here.}.
At this point, the Hubble parameter $H$ will be constant as well.
The cosmological equations become:
\bea
&& \sqrt{H_0^2 + \frac{\L}{10}} + \sqrt{H_0^2 + \s^2 \frac{\L}{10}} = \frac{\l}{4M^4}
- \frac{3}{2} r_c H_0^2  \label{lateeq_1} \\
&& -H_0^2 \left[ 4  \left(H_0^2 + \frac{\L}{10} \right)^{-\frac{1}{2}} +
4 \left(H_0^2 + \s^2\frac{\L}{10} \right)^{-\frac{1}{2}} +
3 \frac{\sqrt{H_0^2 + \s^2 \frac{\L}{10}}} {H_0^2 + \frac{\L}{10}} \right] - \nn \\
&& -\frac{\L}{10} \left[ 4\left(H_0^2 + \frac{\L}{10} \right)^{-\frac{1}{2}} +
\s^2 \left(H_0^2 + \s^2 \frac{\L}{10} \right)^{-\frac{1}{2}} +
3 \frac{\sqrt{H_0^2 + \s^2 \frac{\L}{10}}}{H_0^2 + \frac{\L}{10}} \right] = \nn \\
&& = -\frac{\l}{M^4} + 12 r_c H_0^2 \label{lateeq_2}
\eea
with the conservation equation trivially satisfied. These are a set of two algebraic equations
in the unknown $H_0$, $\L$ and $\l$, whose solution\footnote{Again, since the system can be
cast into a set of two fourth-order equations in $H_0^2$, an analytical solution
could be found, but its (very complicated) form is unimportant here.}  gives the
curvature in terms of the bulk cosmological constant and the brane tension,
and a unavoidable fine-tuning between these last two parameters. A similar situation occurs
if we set $w = -1$, so that the energy density just results in a further contribution to the
brane tension. From eq. (\ref{compeq_2}) we see that the extradimensional pressure $P$ is also
constant, and can be expressed in terms of the energy density. Then we eventually get eqs.
(\ref{lateeq_1},\ref{lateeq_2}) again, with a rescaled brane tension.

On the other side, we can assume that the energy density dominates over cosmological
constant and tension, so that we can ignore the latter two.
In addition, since the Hubble parameter has to be the same order of magnitude
as $\rho$, we can also ignore cosmological constant when added to $H^2$.
In this way eqs. (\ref{compeq_1}),(\ref{compeq_2}),(\ref{compeq_3}) greatly simplify.
We then substitute the extradimensional pressure $P$ as evaluated from (\ref{compeq_3})
in (\ref{compeq_2}), so to obtained the modified Friedman equations
(from now on we substitute $\rho/M^4 \rarr \rho$, $P/M^4 \rarr P$
and $\tau \rarr t$):
\bea
&& 2H = \frac{\rho}{4} - \frac{3}{2}r_c H^2 \label{denseq_1} \\
&& \dot{\rho} + H \left[ \left(4+3w \right)\rho + 3r_c \left(\dot{H}+4H^2 \right) \right]
+ 2\dot{H} + 11H^2 \label{denseq_2}
\eea
Let us stress that the energy density we are considering in these equation is
a 5-dimensional energy density. The observable 4-dimensional energy density is obtained
by integrating over the compact direction; again, in the limit we are considering
$a(\tau) \gg \mu $ it is easy to see that:
\be
\4 \rho \propto \frac{a}{\L}\rho
\label{4-D_rho_late}
\ee

Equations (\ref{denseq_1}),(\ref{denseq_2}) can be further approximated by noting that the Hubble
radius can be either much greater or much smaller than the crossover scale $r_c$, so we will
consider the two cases separately:
\begin{itemize}
\item{\it Sub-crossing regime}:

In this case we have $r_c H \gg 1$, so that we can discard terms not proportional to $r_c$.
The equations (\ref{denseq_1}),(\ref{denseq_2}) can be exactly solved to give:
\bea
a(t) &=& \left( \frac{t}{t_0}\right)^{\frac{5}{6(w+2)}} \label{a_sub_sol} \\
\rho(t) &=& \rho_0 \left( a(t) \right)^{-\frac{6}{5}(w+2)} \label{rho_sub_sol}
\eea
Then we can use (\ref{4-D_rho}) to express the behavior of $\4 \rho$ with respect
to the scale factor, specializing the results for the case of interest of radiation ($w=1/3$)
and matter ($w = 0$). We have:
\bea
\4 \rho_r &=& \4 \rho_{r,0}a^{-\frac{9}{5}} \label{subhor_rad} \\
\4 \rho_m &=& \4 \rho_{m,0}a^{-\frac{7}{5}} \label{subhor_mat}
\eea

\item{\it Super-crossing regime}:

In this case we have $r_c H \ll 1$ so we can drop terms proportional to $r_c$ in
(\ref{denseq_1}),(\ref{denseq_2}). The solutions are:
\bea
a(t) &=& \left( \frac{t}{t_0}\right)^{\frac{10}{24w+43}} \label{a_super_sol} \\
\rho(t) &=& \rho_0 \left( a(t) \right)^{-\frac{24w+43}{10}} \label{rho_super_sol}
\eea
which become, for the 4-dimensional radiation and matter energy density:
\bea
\4 \rho_r &=& \4 \rho_{r,0}a^{-\frac{41}{10}} \label{superhor_rad} \\
\4 \rho_m &=& \4 \rho_{m,0}a^{-\frac{33}{10}} \label{superhor_mat}
\eea
\end{itemize}

To analyze the behavior of the brane at early times, we need to revert the approximation
we made at the beginning of this section, and assume $a(t) \ll \mu$. After some algebra,
substituting P from (\ref{compeq_3}) in (\ref{compeq_2}) as before,
the modified Friedman equations can be approximated as:
\bea
&& 2H = 2\frac{a}{\mu}\rho - 3r_c H^2 \label{early_eq_1} \\
&& \dot{\rho} +\frac{\mu H}{2a} \left[ \rho + \frac{3}{2}r_c
\left( \dot{H} + 4H^2\right) - \frac{\mu H}{a} \right]= 0
\label{early_eq_2} \eea

\begin{figure}[ht]
\begin{center}
\epsfig{file=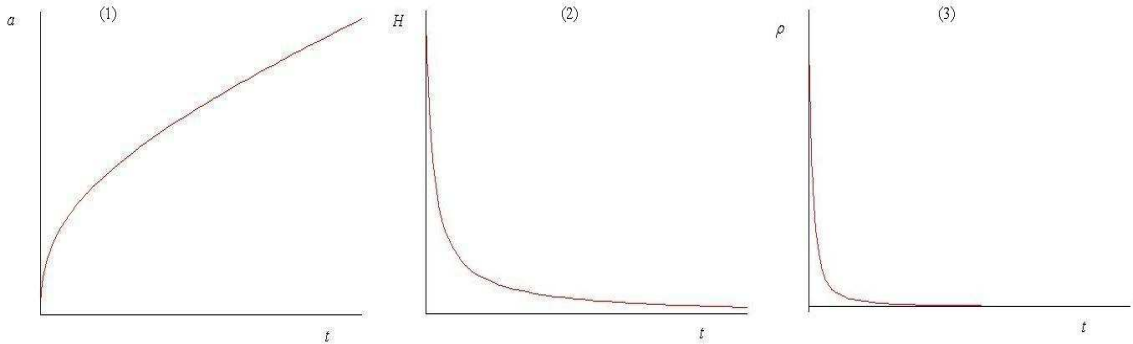,width=16cm,height=5cm} \caption{Plots of
the scale factor $a$ (1), the Hubble parameter $H$ (2) and the
4-dimensional energy density $\4 \rho$ (3) as obtained from
numerical solutions of eqs.
(\ref{early_eq_1}),(\ref{early_eq_2}).} \label{early_plots}
\end{center}
\end{figure}

These equations cannot be solved analytically, so we must turn to numeric. The behavior
of the scale factor $a(t)$, the Hubble parameter $H(t)$ and the 4-dimensional energy
density $\4 \rho(t)$ are shown in Fig. \ref{early_plots}. Notice that, with the approximation
$a(t) \ll \mu$ we are using, the 4-dimensional energy density is related with the 5-dimensional
one by the relation:
\be
\4 \rho \propto \frac{\sqrt{\mu a}}{\L}\rho
\label{4-D_rho_early}
\ee

\section{Comments and conclusions}
\label{comm_concl}

In this paper we have studied the cosmological properties of a 5-dimensional brane,
described by the action (\ref{action}), assumed to be a regularization of a codimension-2
braneworld model. Cosmological evolution is governed by the movement of the brane
trough the extra-dimension, while 4-dimensional cosmology is obtained by
integrating over the fifth compact dimension. The cosmological equations
can be specialized to describe different regimes.

Close to the inner cap, evolution of the brane
seem to emerge from an initial singularity, much alike what happens in standard cosmology.
It is known that extradimensional contribution to the energy-momentum tensor result in an effective
negative energy density \cite{Vollick:2000uf}, and codimension-1 models have been proposed
\cite{Mukherji:2002ft,DeRisi:2007dn} in which this contribution dominates at early times, thus
providing a non-singular brane cosmology. Eqs. (\ref{early_eq_1}), (\ref{early_eq_2})
shows that this negative contribution is actually present in our model, but
is exactly cancelled by the modified dynamics of the energy density, so that even if the static
space-time is non-singular in the origin, still cosmological dynamics is plagued by
an initial singularity (the 4-dimensional curvature is roughly $\4 R \propto H^2/a^{3/2}$).

Next, the universe is supposed to enter in an energy-dominated phase. In the present model
we assume that different sources dominates at different eras, so single contributions
can be considered independently. The presence of a curvature term on the brane
indicates that a crossover scale $r_c$ can be identified, which is given by the ratio
between 5-dimensional and 6-dimensional gravitational coupling constants, and which
should represent the scale at which extradimensional physics become effective.
In the DGP scenario \cite{Dvali:2000hr}, these extradimensional effects can provide,
at super-crossing scale, for a self-accelerated
expansion without the necessity of a cosmological constant \cite{Deffayet:2000uy}.
Here the dynamics is quite different. We find that, if energy density dominates at
a scale smaller than the crossover scale, cosmology on the brane differs evidently from
standard 4-dimensional cosmology. If, on the contrary, radiation and matter eras begin on a
large enough scale (or, equivalently, if the crossover scale is small enough), we find
(\ref{superhor_rad}), (\ref{superhor_rad}) that energy density scales with respect
to scale factor with almost the same power-law as standard cosmology,
differently for what happens in other examples of induced cosmology on a codimension-2 brane
 presented in the literature.

Eventually, energy density will drop below the order of magnitude of the
cosmological constant and the tension (notice that the model under discussion depends
crucially on the presence of a bulk cosmological constant to be dynamically meaningful);
at that point, the brane undergoes a phase of
de Sitter expansion, with an effective cosmological constant given by the solution
of eqs. (\ref{lateeq_1}), (\ref{lateeq_2}). These equations impose also an unavoidable
fine tuning between bulk cosmological constant and brane tension. It is possible
that such a dependence could arise in the process of nucleation of the brane, which is
of course fully nonlinear and very difficult to describe. A hint towards this assumption
comes from the observation that the effective brane tension can be rescaled by tuning
the extradimensional pressure. Still, even at the level of
our analysis, it seems that the self-tuning property is lost once we go beyond the static
solutions, since the effective cosmological constant on the brane has a non-trivial
dependence on the tension.

Summing up, our investigation seems to suggest that a braneworld model
embedded in a conical de Sitter bulk could have a viable cosmology, i.e.
the evolution of the brane during radiation and matter dominated phase is
similar to what happens in standard cosmology, but, at the level of our analysis,
the self-tuning property is lost; neither extradimensional contribution
are enough to address the initial singularity problem.

In order to address these drawbacks, it would be important to develop a model
in which brane movement could modify bulk geometry, so that the deficit angle could
become in some sense ``dynamical''. This is a formidable problem, as said elsewhere,
because it would require a solution of the full 6-dimensional problem, which means
tackling a set of non-linear partial differential equations. We are working in this
topic, though preliminary results are not encouraging. A question remains on how
reliable is a ``mirage cosmology'' approximation when applied to a codimension-2 cosmological brane
(though regularized). We assume that the probe brane approximation should work as long as
the curvature do not blows up. In this spirit, the undesirable singularity should not
be taken too seriously, also because in a realistic model the brane should emerge from a nucleation
process in the bulk. However, the lack of self-tuning should be a robust prediction,if some other
effect do not change the picture drastically (such as supersymmetry).

Another important development would be studying of perturbations around the background we have presented.
Perturbations around the static solution give, as stated in section \ref{intro},
the tensorial structure of 4-dimensional gravity, which has an unwanted scalar degree of
freedom that propagate on the brane. It is conjectured that this degree of freedom
wold be reabsorbed in a nonlinear realization of our codimension-2 model. Perturbations
would allow us to identify the exact form of the tree-level graviton exchange between two
probe masses in 4-dimensional gravity, which, since 4-dimensional physics is obtained by integrating
a time-varying azimuthal dimension, would imply a time-varying Planck mass and observational signatures
that would change in different cosmological epoch as well. This could results in
a tight constraint on the parameters of the model (and of other regularized codimension-2 brane models)
from space-based experiments. Unfortunately, it is not clear how to obtain reliable perturbation equations
starting from mirage cosmology; so in order to perform these very interesting investigation, one is again led to
the necessity of a full 6-dimensional nonlinear study of brane cosmology.

 A simpler, and yet interesting development would be to study the
supersymmetric extension of the present model, which could help to improve the fine-tuning
problem between the brane tension and the bulk cosmological constant with a suitable
dilaton potential. All these aspects will be addressed in forthcoming works.

\section*{Acknowledgements}
It is a pleasure to thank Antonio Cardoso, Olindo Corradini, Maurizio Gasperini, Kazuya Koyama,
Roy Maartens, Antonios Papazoglou, Fabio Silva and David Wands for helpful discussion and
comments on the manuscript

\end{document}